\documentclass[%
 aip,
 amsmath,amssymb,
 reprint,%
]{revtex4-1}

\usepackage[english]{babel}
\usepackage{graphicx}
\usepackage{bm}
\usepackage[utf8]{inputenc}
\usepackage[T1]{fontenc}
\usepackage{mathptmx}
\usepackage{booktabs}
\usepackage{etoolbox}
\usepackage{physics}
\usepackage[colorlinks=true,
            linkcolor=blue,
            urlcolor=blue,
            citecolor=blue]{hyperref}

\DeclareMathOperator{\cov}{cov}

\makeatletter
\def\@email#1#2{%
 \endgroup
 \patchcmd{\titleblock@produce}
  {\frontmatter@RRAPformat}
  {\frontmatter@RRAPformat{\produce@RRAP{*#1\href{mailto:#2}{#2}}}\frontmatter@RRAPformat}
  {}{}
}%
\makeatother
\begin{document}

\preprint{AIP/123-QED}

\title[Stickiness and recurrence plots: an entropy-based approach]{Stickiness and recurrence plots: an entropy-based approach}
\author{Matheus R. Sales}
\email{matheusrolim95@gmail.com}
\affiliation{Graduate Program in Sciences/Physics, State University of Ponta Grossa, 84030-900, Ponta Grossa, PR, Brazil}
\affiliation{Potsdam Institute for Climate Impact Research, Member of the Leibniz Association, P.O. Box 6012 03, D-14412 Potsdam, Germany}
\affiliation{Institute of Mathematics, Humboldt University Berlin, 12489 Berlin, Germany}
\author{Michele Mugnaine}
\affiliation{Department of Physics, Federal University of Paraná, 80060-000, Curitiba, PR, Brazil}
\affiliation{Institute of Physics, University of S\~ao Paulo, 05508-900, S\~ao Paulo, SP, Brazil}
\author{José D. Szezech Jr.}
\affiliation{Graduate Program in Sciences/Physics, State University of Ponta Grossa, 84030-900, Ponta Grossa, PR, Brazil}
\affiliation{Department of Mathematics and Statistics, State University of Ponta Grossa, 84030-900, Ponta Grossa, PR, Brazil}
\author{Ricardo L. Viana}
\affiliation{Department of Physics, Federal University of Paraná, 80060-000, Curitiba, PR, Brazil}
\affiliation{Institute of Physics, University of S\~ao Paulo, 05508-900, S\~ao Paulo, SP, Brazil}
\author{Iber\^e L. Caldas}
\affiliation{Institute of Physics, University of S\~ao Paulo, 05508-900, S\~ao Paulo, SP, Brazil}
\author{Norbert Marwan}
\affiliation{Potsdam Institute for Climate Impact Research, Member of the Leibniz Association, P.O. Box 6012 03, D-14412 Potsdam, Germany}
\author{Jürgen Kurths}
\affiliation{Potsdam Institute for Climate Impact Research, Member of the Leibniz Association, P.O. Box 6012 03, D-14412 Potsdam, Germany}
\affiliation{Institute of Physics, Humboldt University Berlin, 10099 Berlin, Germany}
\affiliation{Division of Dynamics, Lodz University of Technology, Stefanowskiego 1/15, 90-924 Lodz, Poland}

\date{\today}

\begin{abstract}
    The stickiness effect is a fundamental feature of quasi-integrable Hamiltonian systems. We propose the use of an entropy-based measure of the recurrence plots (RP), namely, the entropy of the distribution of the recurrence times (estimated from the RP), to characterize the dynamics of a typical quasi-integrable Hamiltonian system with coexisting regular and chaotic regions. We show that the recurrence time entropy (RTE) is positively correlated to the largest Lyapunov exponent, with a high correlation coefficient. We obtain a multi-modal distribution of the finite-time RTE and find that each mode corresponds to the motion around islands of different hierarchical levels.
\end{abstract}
\keywords{recurrence plots, quasi-integrable Hamiltonian system, stickiness, recurrence time entropy}
\maketitle

\begin{quotation}
In two-dimensional quasi-integrable Hamiltonian systems with hierarchical phase space, chaotic orbits can spend an arbitrarily long time around islands, in which they behave similarly as quasiperiodic orbits. This phenomenon is called stickiness, and it is due to the presence of partial barriers to the transport around the hierarchical levels of islands-around-islands. The stickiness affects the convergence of the Lyapunov exponents, making the task of characterizing the dynamics more difficult, especially when only short time series are known. Due to the intrinsic property of dynamical systems that quasiperiodic orbits can have at most three different return times (Slater's theorem \cite{slater_1950, slater1967}), which is the time needed to the orbit return to a given region at the curve, in this paper we propose the use of the recurrence time entropy (RTE) (estimated from the recurrence plots) to characterize the dynamics of nonlinear systems. We find that the RTE is an alternative way of detecting chaotic orbits and sticky regions. Furthermore, the finite-time RTE distribution is multi-modal when sticky regions are present in the phase space, and each mode corresponds to a different hierarchical level in the islands-around-islands structure embedded in the chaotic sea.

\end{quotation}

\section{Introduction}
\label{sec:intro}

The phase space of a typical quasi-integrable Hamiltonian system is in general neither integrable nor uniformly hyperbolic, but there is a coexistence of chaotic and regular domains \cite{lichtenberg2013regular}. The regular dynamics consists of periodic and quasiperiodic orbits that lie on invariant tori, whilst the chaotic orbits fill densely the available domain in phase space. For the special case of two-dimensional area-preserving maps, the existence of islands of regularity filled with Kolmogorov-Arnold-Moser (KAM) invariant tori, separates the phase space into distinct regions, \textit{i.e.}, orbits in the chaotic sea will never enter any island, and the periodic and quasiperiodic orbits inside of an island will never reach the chaotic sea \cite{cantori2, lichtenberg2013regular}. Due to the existence of islands embedded in the chaotic sea, the latter constitutes a fat fractal \cite{fat_fractal} and it is challenging to determine exactly the islands' boundary. The islands are surrounded by smaller islands, which are in turn surrounded by even smaller islands. This structure repeats itself for arbitrarily small scales giving rise to an infinite hierarchical islands-around-islands structure \cite{meiss_transport}.

The phenomenon of stickiness \cite{stickContopoulos,stickKarney, meissdecay, stickinesscantori, PhysRevLett.100.184101} emerges due to this complex interplay between islands and chaotic sea. The chaotic orbits that approach an island may spend an arbitrarily long, but finite, time in its neighborhood, in which the orbits will behave similarly as quasiperiodic orbits. Before escaping, the orbits are trapped in a region bounded by cantori \cite{cantori1, cantori2,stickinesscantori}, which are a Cantor set, formed by the remnants of the destroyed KAM tori. Unlike the KAM tori that are full barriers to the transport in phase space, the cantori act as partial barriers, where the orbits may be trapped in the region bounded by them, and once trapped inside a cantorus, the chaotic orbits may cross an inner cantorus, and so on to arbitrarily small levels in the hierarchical structure of islands-around-islands.

Since in two-dimensional area-preserving maps the islands and the chaotic sea are distinct and disconnected domains, it is of major importance to characterize orbits into these two categories. The traditional and most known method to characterize the dynamics of a system is through the evaluation of the Lyapunov exponents \cite{WOLF1985285,eckmann}. For a two-dimensional mapping, there are two Lyapunov exponents, $\lambda_1$ and $\lambda_2$, and the dynamics is chaotic if one of them is positive. When the mapping is area-preserving, the sum of the two exponents must be zero, \textit{i.e.}, $\lambda_1 = -\lambda_2$. In this scenario, the regular orbits have Lyapunov exponents equal to zero for infinite times, while the chaotic ones exhibit $\lambda_1 > 0$. 

If sticky regions are present in the phase space, the Lyapunov exponents may not be the optimal choice to detect chaotic orbits due to the trappings around the islands. When the orbit is trapped, the largest Lyapunov exponent decreases and this makes its convergence slower, \textit{i.e.}, it takes longer to reach the asymptotic (infinite-time) value. As an alternative, a new method based on ergodic theory has recently been proposed to detect chaotic orbits \cite{Das_2016,SANDER2020132569,MEISS2021133048,SALES2022127991}. It proved to be a better option to distinguish between regular and chaotic orbits than the Lyapunov exponents. However, both this new method and the Lyapunov exponents require very long time series in order to obtain reliable accuracy. When only a short time series is available, a possible approach is to use the recurrence quantification analysis (RQA) \cite{rec1, rec2, rec3, rec4, rec5}. The RQA was developed to quantify the dynamics of a system by means of the recurrences of the orbit in phase space. 


The most used RQA measures (\textit{e.g.} the recurrence rate and the determinism) can, in some sense, detect different transitions occurring in nonlinear systems. However, we seek a measure based on an intrinsic property of dynamical systems: quasiperiodic orbits lying on invariant circles can have at most three different return (recurrence) times, which is the time needed for the orbit return to a given neighborhood of a point in the orbit, as stated by Slater's theorem \cite{slater_1950, slater1967}. We can obtain the recurrence times by simply defining a recurrence region, and counting how long it takes to the orbit to return to this given region. It is also possible to use the recurrence plots (RPs) to estimate the recurrence times: the white vertical lines in the RP give us a lower estimate of the recurrence times \cite{zou_sticky, zou2007phd, zou2, entropy3, PhysRevE.85.026217}. Thus, in this paper, we propose using RPs to characterize the dynamics of a nonlinear system. We focus on a further type of RP-based measure to identify regular and chaotic regions, and sticky regions as well, namely the Shannon entropy of the recurrence times: the recurrence time entropy (RTE) \cite{entropy2,RTE}. The RTE was originally introduced without any connection to the RPs \cite{entropy2}, and it was shown that it can provide a good estimate for the Kolmogorov-Sinai entropy \cite{entropy3} and the largest Lyapunov exponent \cite{shiozawa}.

We find that, with the RTE, we can identify very clearly the regular regions and the transitions to chaotic motion as one parameter of the system is varied. We also find that, by computing the finite-time RTE distribution, we can identify a multi-modal distribution, in which each maximum is related to a different hierarchical level in the islands-around-islands structure. Moreover, each of these regions corresponds to a power-law decay of the cumulative distribution of trapping times.

The paper is organized as follows. In Section \ref{sec:stdmap} we introduce the standard map and briefly comment on some of the properties of two-dimensional quasi-integrable Hamiltonian systems. We also discuss a few approaches of how to detect sticky orbits in the phase space of such systems. In Section \ref{sec:recplot} we introduce the concept of recurrence plots and show that it can be used to characterize the dynamics of a dynamical system. In this Section, motivated by Slater's theorem, we also propose the use of the RTE, estimated from the RP, to quantify the dynamics as periodic, quasiperiodic, and chaotic. In Section \ref{sec:wvle} we apply this entropy-based measure for the standard map and show that it is positively correlated to the largest Lyapunov exponent, with a high correlation coefficient, and also show that it is possible to detect and characterize different regimes of stickiness present in the dynamics. Section \ref{sec:conclusion} contains our final remarks.

\section{The standard map}
\label{sec:stdmap}

The standard map \cite{chirikovstdmap}, also known as Chirikov-Taylor map, is a two-dimensional area-preserving map and its dynamics is given by the following equations
\begin{equation}
    \label{eq:stdmap}
    \begin{aligned}
        x_{n + 1} &= x_n + p_{n + 1}\mod{2\pi},\\
        p_{n + 1} &= p_n - k\sin{x_n}\mod{2\pi},
    \end{aligned}
\end{equation}
where $x_n$ and $p_n$ are the canonical position and momentum, respectively, at discrete times $n=0,1,2,\ldots,N$ and $k$ is the nonlinearity parameter.

In spite of its simple mathematical form, the standard map exhibits all the features of a typical quasi-integrable Hamiltonian system, and it has become a paradigmatic model for the study of properties of chaotic motion in quasi-integrable Hamiltonian systems. For $k = 0$ the dynamics is regular, the system is integrable and all orbits lie on invariant rotational tori. As the nonlinearity parameter $k$ increases, the ``sufficiently irrational'' invariant tori persist, as predicted by the KAM theorem \cite{lichtenberg2013regular}, whereas the rational tori are destroyed. When $k$ is sufficiently large, it turns out that all the invariant rotational tori are destroyed. For the standard map, the last invariant rotational torus ceases to exist for the critical value $k \approx 0.971635$ \cite{greene}, leading to a scenario of global stochasticity.

\begin{figure}[tb]
    \centering
    \includegraphics[width=0.9\linewidth]{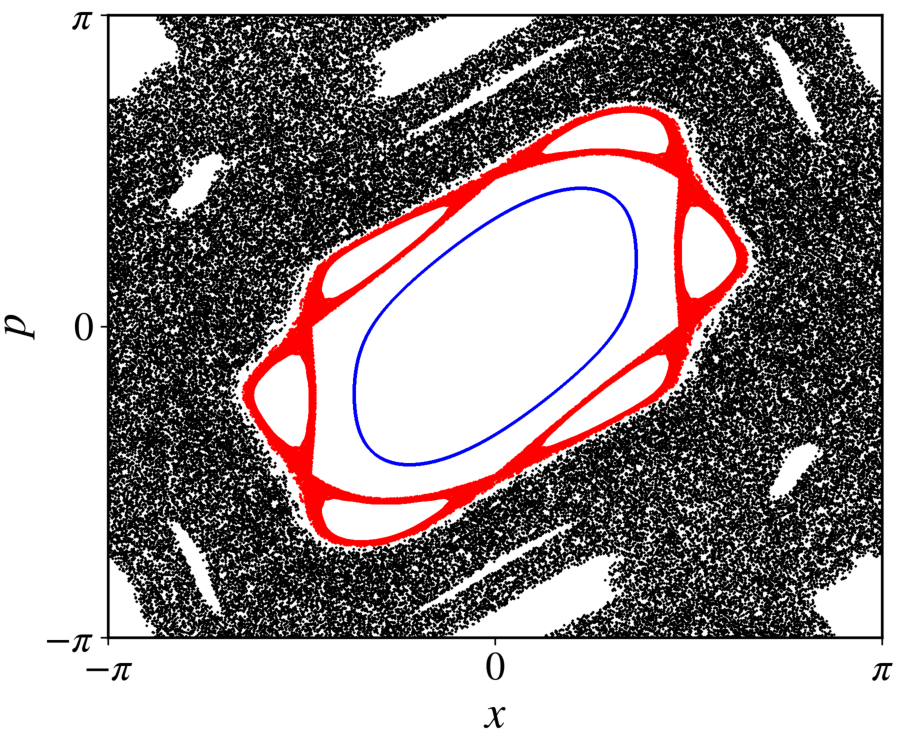}
    \caption{Phase space of a quasiperiodic orbit (blue), chaotic orbit (black) and sticky orbit (red) of the standard map \eqref{eq:stdmap} with $k = 1.5$ iterated for $N = 8.5\times10^4$ times. The largest Lyapunov exponent of each orbit is $\lambda_{\mathrm{max}} = 0.00017$, $\lambda_{\mathrm{max}} = 0.42896$ and $\lambda_{\mathrm{max}} = 0.30666$, respectively.}
    \label{fig:phase_space}
\end{figure}

One of the main features of quasi-integrable Hamiltonian systems is the phenomenon of stickiness. In Figure \ref{fig:phase_space} are displayed the phase space of a quasiperiodic orbit (blue), a chaotic orbit (black), and a sticky orbit (red) of the standard map \eqref{eq:stdmap} with $k = 1.5$ and the initial conditions $(x_0, p_0) = (1.0, 0.0)$, $(x_0, p_0) = (2.9, 0.0)$ and $(x_0, p_0) = (1.6, 0.0)$, respectively. The orbits were iterated for $N = 8.5 \times 10^4$ times and we also calculated the largest Lyapunov exponent, $\lambda_{\mathrm{max}}$, of each orbit, namely, $\lambda_{\mathrm{max}} = 0.00017$, $\lambda_{\mathrm{max}} = 0.42986$ and $\lambda_{\mathrm{max}} = 0.30666$, respectively. For this value of $k$, one single chaotic orbit fills a significant portion of phase space (black dots) and the most prominent sticky region is around the period-6 satellite islands (red dots). An orbit initialized in this region is trapped for a long time until it escapes to the chaotic sea, and this affects some properties of the orbit, the $\lambda_{\mathrm{max}}$ in particular. Although chaotic and sticky orbits both have $\lambda_{\mathrm{max}} > 0$, the sticky orbit has a lower value of $\lambda_{\mathrm{max}}$ than the chaotic one. The $\lambda_{\mathrm{max}}$ for the quasiperiodic orbit is small, but not exactly zero due to the finite iteration time $N$: as $N \rightarrow \infty$, $\lambda_{\mathrm{max}} \rightarrow 0$. Furthermore, the closer the quasiperiodic orbit is to the elliptic point, the faster the convergence of $\lambda_{\mathrm{max}}$ towards zero \cite{cgbd}.

The stickiness is usually characterized through the recurrence time statistics \cite{GRASSBERGER1985167,meiss_transport,PhysRevLett.82.528,ZASLAVSKY2002461,PhysRevE.67.046209,PhysRevLett.100.184101, crt1, crt2}, although other methods have been proposed, such as the finite-time Lyapunov exponent (FTLE) \cite{ftle,HARLE2007130,PhysRevE.91.062907,PhysRevE.99.052208}, the rotation number \cite{rotation_number_sticky}, and the recurrence quantification analysis (RQA) \cite{zou_sticky, zou2007phd, zou2, palmero}. Szezech \textit{et al.} \cite{ftle} showed that for the case where the phase space has stickiness regions, the distribution of the finite-time Lyapunov exponents is bimodal. The finite-time Lyapunov exponent has also been used to characterize stickiness in high-dimensional Hamiltonian systems \cite{PhysRevE.91.062907,PhysRevE.99.052208}. More recently, Santos \textit{et al.} \cite{rotation_number_sticky} showed that the rotation number is a faster method, compared to the finite-time Lyapunov exponent, to verify the presence of sticky orbits in the phase space. The most commonly used measures of the RQA, such as the determinism and the recurrence rate, can also characterize stickiness in a similar way \cite{zou2007phd,zou_sticky, palmero}.

In the next section, we review the concept of recurrence plots and we propose a quantity based on them, different from the traditional measures of RQA, to characterize the dynamics of a nonlinear system.

\section{Recurrence plots}
\label{sec:recplot}

\begin{figure*}[tb]
    \centering
    \includegraphics[width=0.95\linewidth]{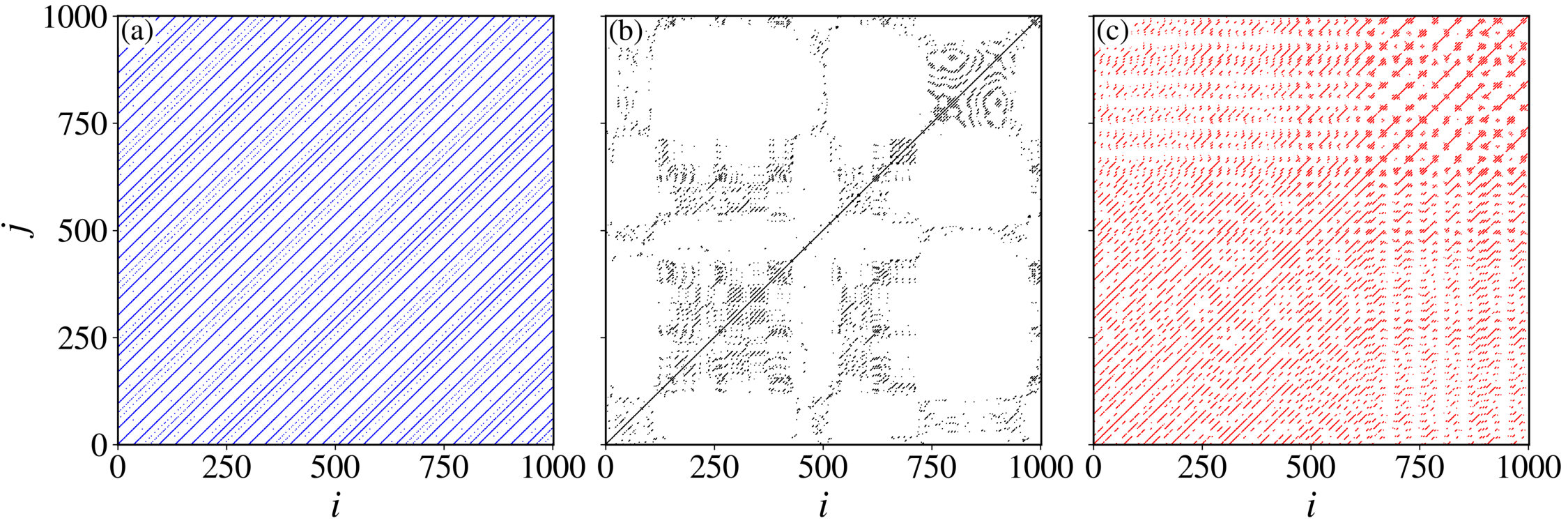}
    \caption{Recurrence matrix of the (a) quasiperiodic orbit, (b) chaotic orbit and (c) sticky orbit of the standard map \eqref{eq:stdmap} with $k = 1.5$ shown in Figure \ref{fig:phase_space}.}
    \label{fig:recmats}
\end{figure*}

The recurrence plot (RP), first introduced by Eckmann \textit{et al.} in 1987 \cite{Eckmann_1987}, is a graphical representation of the recurrences of time series of dynamical systems in its $d$-dimensional phase space. Given a trajectory $\vec{x}_i \in \mathbb{R}^d$ ($i = 1, 2, \ldots, N)$, we define the $N\times N$ recurrence matrix as
\begin{equation}
    \label{eq:recmat}
    R_{ij} = \Theta\qty(\epsilon - \|\vec{x}_i - \vec{x}_j\|),
\end{equation}
where $i, j = 1, 2, \ldots, N$, ($N$ is the length of the time series), $\Theta$ is the Heaviside unit step function, $\epsilon$ is a small threshold, and $\|\vec{x}_i - \vec{x}_j\|$ is the spatial distance between two states, $\vec{x}_i$ and $\vec{x}_j$, in phase space in terms of a suitable norm. In this work we consider the supremum (or maximum) norm.

The recurrence matrix $\vb{R}$ is a symmetric, binary matrix that contains the value $1$ for recurrent states and the value $0$ for nonrecurrent ones. Two states are said to be recurrent when the state at $t = i$ is ``close'' (up to a distance $\epsilon$) to a different state at $t = j$, \textit{i.e.}, $\vec{x}_i\approx\vec{x}_j$. The choice of the threshold $\epsilon$ is not arbitrary. If $\epsilon$ is chosen too large, almost every point is recurrent with every other point. On the other hand, if $\epsilon$ is chosen too small, there will be almost no recurrent states. Several rules have been proposed. Some consider $\epsilon$ with a fixed recurrence point density of the RP \cite{defeps3}. Another possibility is to consider $\epsilon$ as a fraction of the standard deviation, $\sigma$, of the time series \cite{defeps4, Schinkel2008}. Regardless of the choice we make, the effect of a finite $\epsilon$ will never disappear; a new study has shown that using $\epsilon \rightarrow 0$ is not the best choice \cite{doi:10.1063/5.0055797} and in this work we consider the threshold to be $10\%$ of the time series standard deviation, considering the maximum norm approach (see Appendix \ref{sec:appendix}), \textit{i.e.}, $\epsilon = \sigma/10$. For a detailed discussion about the effect of $\epsilon$ in our results, see Appendix \ref{sec:appendix}.

Graphically, the recurrent states are represented by a colored dot, and the recurrence matrix $\vb{R}$ displays different patterns according to the dynamics of the underlying system. In Figure \ref{fig:recmats} we show the recurrence matrix for the first $1000$ iterations of (a) a quasiperiodic orbit, (b) a chaotic orbit, and (c) a sticky orbit of the standard map with $k = 1.5$ shown in Figure \ref{fig:phase_space}. The RP of the quasiperiodic orbit consists mainly of uninterrupted diagonal lines [Figure \ref{fig:recmats}(a)]. The vertical distance between these lines is regular and corresponds to the different return times of the orbit \cite{zou2007phd,zou_sticky, zou2, PhysRevE.85.026217}, whereas the RP of the chaotic orbit displays short diagonal lines and the vertical distances between them are not as regular as in the quasiperiodic case. The RP of the sticky orbit seems to be between these two cases. The diagonal lines are longer than those in the chaotic case, indicating that the sticky orbit is more regular than the chaotic one, but the diagonal lines are not as long as those in the quasiperiodic case. Moreover, the vertical distances between the diagonal lines have some regularity.

Therefore, we can use the RP to quantify the dynamics of the system. Several measures based on the length $\ell$ of the diagonal lines and based on the length $v$ of the vertical lines have been proposed, such as the recurrence rate, the determinism, the laminarity, the maximal length of diagonal and vertical lines, among others. A complete discussion about these and other quantifiers can be found in Refs. \cite{rec1, rec2, rec3, rec4, rec5} and references therein.

Entropy-based quantifiers of RPs have been employed to detect chaotic regimes and bifurcation points \cite{entropy0,entropy1, entropy2, entropy3, entropy4, weightedentropy}. The Shannon entropy of the lines is defined as
\begin{equation}
    \label{eq:shannon}
    S = -\sum_{\ell=\ell_{\mathrm{min}}}^{\ell_{\mathrm{max}}} p(\ell)\ln p(\ell),
\end{equation}
where $\ell_{\mathrm{max}}$ ($\ell_{\mathrm{min}}$) is the length of the longest (shortest) line, $p(\ell) = P(\ell)/N_{\ell}$ and $P(\ell)$ are the relative distribution and the total number of line segments with length $\ell$, respectively, and $N_\ell$ is the total number of line segments. In order to define an RP-based entropy measure based on an intrinsic property of dynamical systems, we recall Slater's theorem \cite{slater_1950,slater1967,mayer1988distribution}. The theorem states that for any irrational linear rotation, with rotation number $\omega$ \footnote{The rotation number, or rotation vector in high-dimensional systems, is a measure of the average increase in the angle variable per unit of time. In two-dimensional systems, if $\omega = p/q$, where $p,q$ are natural numbers, the orbit is periodic with period $q$. If $\omega$ is an irrational number, the orbit is quasiperiodic.}, over a unit circle, there are at most three different return times to a connected interval of size $\delta < 1$. Furthermore, the third return time is always the sum of the other two, and two of them are consecutive denominators in the continued fraction expansion of the irrational rotation number $\omega$. With this in mind, we can distinguish between the different kinds of solutions of a nonlinear system by simply counting the number of return times of an orbit. If it is one the orbit is periodic, and if it is equal to three, the orbit is quasiperiodic. If the number of return times is larger than three, then the orbit is chaotic \cite{zou_sticky, zou2007phd, zou2}. This procedure has been employed to detect chaotic and quasiperiodic orbits of the standard map \cite{zou_sticky} and more recently to study the parameter space of a one-dimensional map \cite{PhysRevE.106.034203}. This is an efficient method. However, it is not obvious how to use it to detect sticky orbits in two-dimensional quasi-integrable Hamiltonian systems.


Since the vertical distances between the diagonal lines in an RP are an estimative of the recurrence times of an orbit, we define the Shannon entropy using the white vertical lines of the RP in Eq. \eqref{eq:shannon}. The total number of white vertical lines (recurrence times) of length $v$ is given by the histogram
\begin{equation}
    \label{eq:wvld}
    P_{\textrm{w}}(v) = \sum_{i, j = 1}^{N}R_{i,j}R_{i,j+v}\prod_{k=0}^{v - 1}(1 - R_{i,j+k}),
\end{equation}
such that the RTE is defined as \cite{entropy2, RTE}
\begin{equation}
    \label{eq:entropy}
    \mathrm{RTE} = -\sum_{v=v_{\textrm{min}}}^{v_{\textrm{max}}}p_{\textrm{w}}(v)\ln{p_{\textrm{w}}(v)},
\end{equation}
where $v_{\mathrm{max}}$ ($v_{\mathrm{min}}$) is the length of the longest (shortest) white vertical line, $p_{\textrm{w}}(v) = P_{\textrm{w}}(v)/N_{\textrm{w}}$ and $N_{\textrm{w}}$ is the total number of white vertical line segments. In this paper, we consider $v_{\mathrm{min}} = 1$. Careful attention should be given to the evaluation of \eqref{eq:wvld}. Due to the finite size of an RP, the distribution of white vertical lines might be biased by the border lines, that are cut short by the borders of the RP, thus influencing the RQA measures, such as the RTE \cite{border}. In order to avoid these border effects, we exclude from the distribution the white vertical lines that begin and end at the border of the RP. The thickening of diagonal lines in an RP, caused by tangential motion \cite{rec3, GAO200075}, which occurs when states $\vec{x}_j$ preceding or succeeding a state $\vec{x}_i$ are within the neighborhood of $\vec{x}_i$ (within $\epsilon$), also influences the RQA measures \cite{border}. However, this effect mainly arises in flows, and in our simulations we find that there is virtually no tangential motion in the dynamics of the standard map (not shown), and we only apply the corrections due to border effects in \eqref{eq:wvld}.


In this way, a periodic orbit, which has only one return time (the period itself), will have $\mathrm{RTE} = 0$. A quasiperiodic orbit, which has three return times, will lead to a low value of $\mathrm{RTE}$, whereas a chaotic orbit will be characterized by a high value of $\mathrm{RTE}$. As has been stated, the chaotic orbit that experiences stickiness spends an arbitrarily long time in the neighborhood of an island in which the orbit exhibits similar behavior as a quasiperiodic orbit. This fact can be seen from the recurrence matrix shown in Figure \ref{fig:recmats}(c). Hence, we expect the RTE of sticky orbits to be smaller than the chaotic ones, but higher than it would be for a quasiperiodic orbit.

\section{Recurrence time entropy}
\label{sec:wvle}

\begin{figure}[tb]
    \centering
    \includegraphics[width=0.95\linewidth]{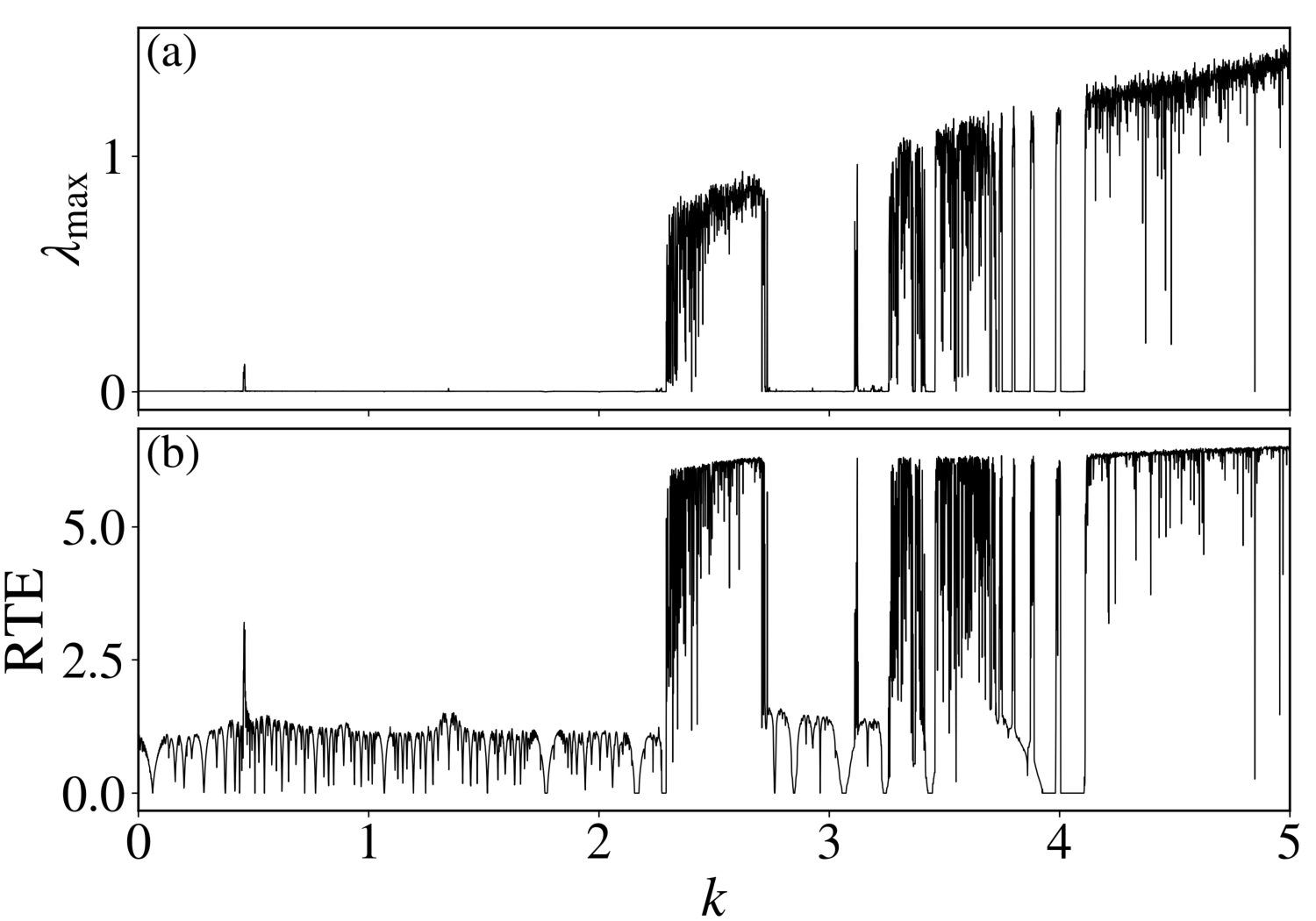}
    \caption{(a) The $\lambda_{\mathrm{max}}$ and (b) the RTE for the standard map \eqref{eq:stdmap} as a function of the parameter $k$ with $x_0 = 0.0$ and $p_0 = 1.3$.}
    \label{fig:vs_k}
\end{figure}
\begin{figure*}[tb]
    \centering
    \includegraphics[width=0.95\linewidth]{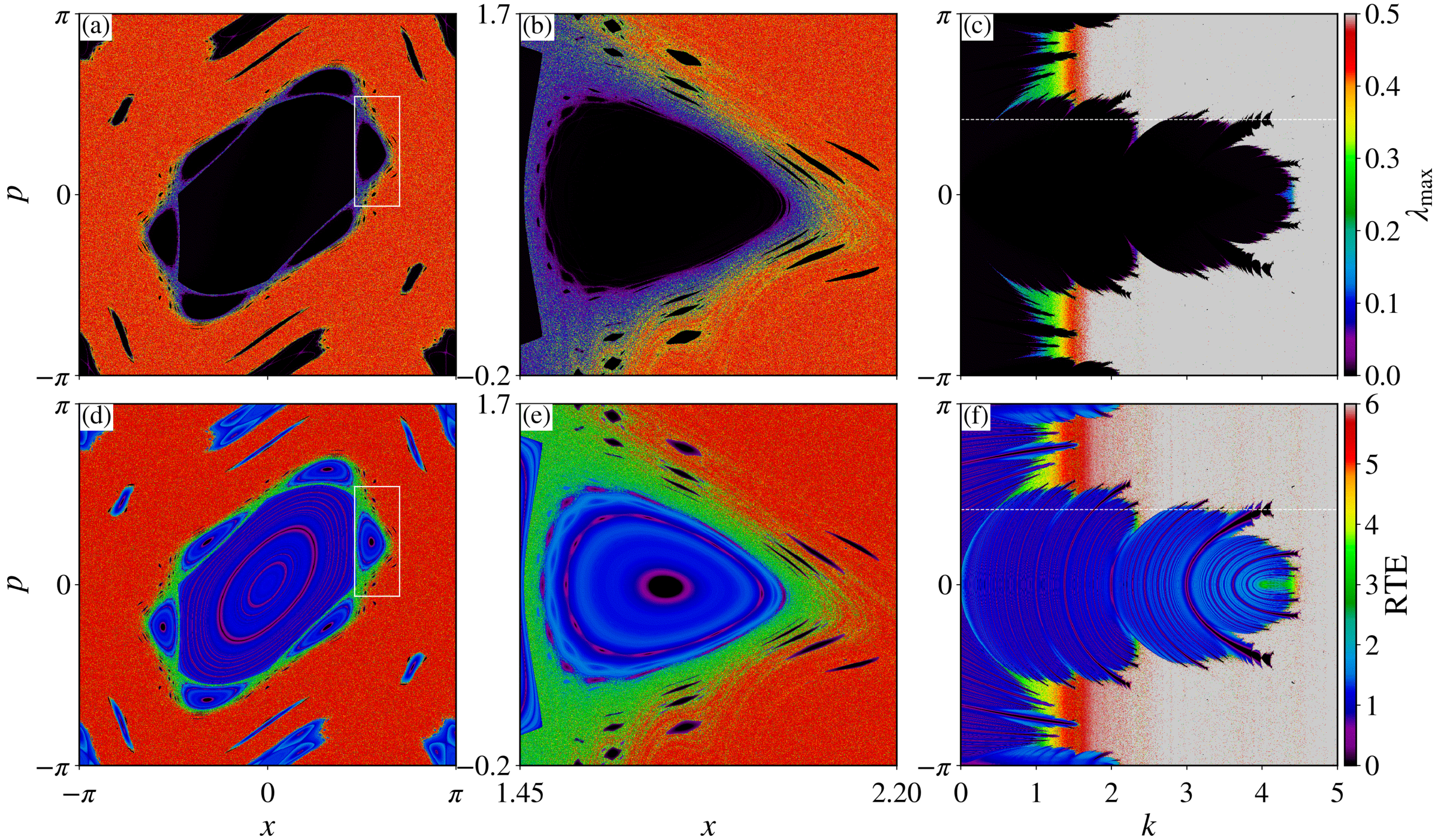}
    \caption{(a)-(c) The $\lambda_{\mathrm{max}}$ and (d)-(f) the RTE for the standard map \eqref{eq:stdmap}, for a $1024\times1024$ grid of uniformly distributed points in the phase space $(x, p$), with $k = 1.5$, for (a), (b), (d) and (e), and in the parameter space $(k, p)$, with $x_0 = 0.0$, for (c) and (f). (b) and (e) are magnifications of the regions bounded by the white rectangle in (a) and (d), respectively, and the dotted white line in (c) and (f) represents the initial condition used in Figure \ref{fig:vs_k}.}
    \label{fig:CGBDs}
\end{figure*}

In this section, we evaluate the $\mathrm{RTE}$ for the standard map, and we show that the RTE can be used to characterize the dynamics of the system and to detect the presence of stickiness regions. Unless mentioned otherwise, we use a time series of size $N = 5000$ for all our simulations.

In Figure \ref{fig:vs_k} we show the $\lambda_{\mathrm{max}}$ and the RTE of the standard map for a fixed initial condition $(x_0, p_0) = (0.0, 1.3)$ as a function of the nonlinearity parameter $k$. We notice that whenever $\lambda_{\mathrm{max}} > 0$, the RTE is large. Also, we can identify the windows of regularity, in which $\lambda_{\mathrm{max}}$ is zero and the RTE assumes low values. Furthermore, even though the Lyapunov exponent goes faster to zero around the elliptic point \cite{cgbd}, only with the values of the Lyapunov exponent we cannot distinguish between the periodic and quasiperiodic orbits in two-dimensional Hamiltonian systems. There are several values of $k$ for which $\mathrm{RTE} \rightarrow 0$, indicating that at these points the initial condition $(x_0, p_0) = (0.0, 1.3)$ is very close to a periodic orbit [Figure \ref{fig:vs_k}(b)].

To have a better visualization of the correspondence between the $\lambda_{\mathrm{max}}$ and the $\mathrm{RTE}$, we plot in Figure \ref{fig:CGBDs} the values of $\lambda_{\mathrm{max}}$ and $\mathrm{RTE}$ for a grid of initial conditions uniformly distributed in the phase space $(x, p)$ with $k = 1.5$, and in the parameter space $(k, p)$ with $x_0 = 0.0$. Figures \ref{fig:CGBDs}(b) and \ref{fig:CGBDs}(e) are a magnification of one of the period-6 satellite islands of Figures \ref{fig:CGBDs}(a) and \ref{fig:CGBDs}(d), respectively. We see that the RTE captures all the features of the Lyapunov exponent but even more. In the chaotic sea, where $\lambda_{\mathrm{max}}$ is large, $\mathrm{RTE}$ is also large and inside the islands, where $\lambda_{\mathrm{max}} \rightarrow 0$, the RTE is low. In addition to that, in the regions where the rotation number of an orbit is close to a rational number, the RTE is smaller (blue to purple) [Figure \ref{fig:CGBDs}(d)] \cite{Das_2016,Das_2018}. The RTE decreases as we get closer to the elliptic point, as we can see in Figures \ref{fig:CGBDs}(e) and \ref{fig:CGBDs}(f). In the latter, we can see the transitions from regular to chaotic behavior, where bifurcations occur as $k$ changes.

For the chosen parameter values of the standard map, $k = 1.5$, it is known that the system exhibits the stickiness effect, and due to that, the distribution of the finite-time Lyapunov exponent is bimodal \cite{ftle}. The most prominent sticky region for this parameter is the region in between the main island and the period-6 satellite islands. The $\lambda_{\mathrm{max}}$ decreases in this region when compared with the rest of the chaotic sea [Figures \ref{fig:CGBDs}(a) and \ref{fig:CGBDs}(b)]. With the RTE we observe the same behavior [Figures \ref{fig:CGBDs}(d) and \ref{fig:CGBDs}(e)].

\begin{figure*}[tb]
    \centering
    \includegraphics[width=0.95\linewidth]{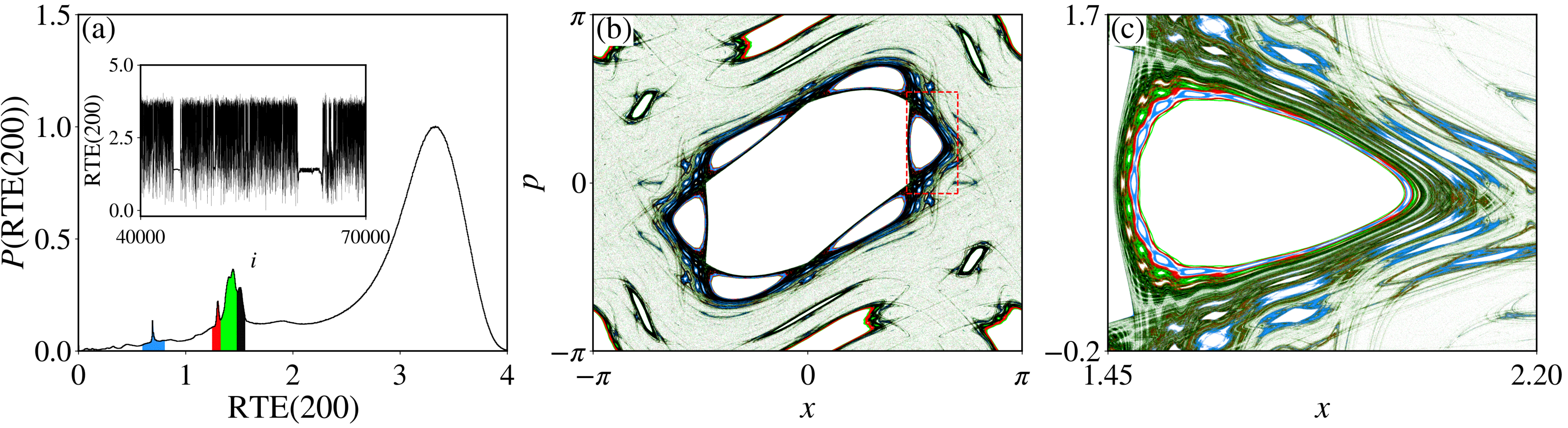}
    \caption{(a) The finite-time RTE distribution for a single chaotic orbit, with $n = 200$, $N = 10^{10}$ and $k = 1.5$, (b) the phase space points that generate the minor peaks in (a) and (c) is a magnification of one of the period-6 satellite islands of (b), indicated by the red dashed lines. The colors in (b) and (c) match the filling colors of (a). Inset: the time series of the finite-time RTE.}
    \label{fig:ftwve}
\end{figure*}

Therefore, at least qualitatively, we can see that the $\mathrm{RTE}$ is positively correlated to $\lambda_{\mathrm{max}}$. In order to quantify this correlation, we use the Pearson correlation coefficient, defined as
\begin{equation}
    \label{eq:correlationcoef}
    \rho_{xy} = \frac{\cov(x, y)}{\sigma_x\sigma_y},
\end{equation}
where $\cov(x,y)$ is the covariance of the two time series, $x$ and $y$, and $\sigma_x$ and $\sigma_y$ are their standard deviation, respectively. Applying \eqref{eq:correlationcoef} to the data in Figures \ref{fig:vs_k} and \ref{fig:CGBDs}, we find that the $\mathrm{RTE}$ is positively correlated to $\lambda_{\mathrm{max}}$ and the value of the correlation coefficients is very close to $1$, indicating a very high correlation (Table \ref{tab:corr}).

Next, we consider a single chaotic orbit of the standard map and we follow up its evolution for a long iteration time $N$. For long times, the trajectory fills the entire chaotic component of the phase space, and usually, the stickiness acts for very long, but finite, times before an orbit escapes to the chaotic sea. Thus, the transitions from different regimes in the dynamics of the orbit can be better understood considering a ``finite-time'' $n \ll N$. Therefore, we compute the RTE along the evolution of a single chaotic orbit in windows of size $n$, $\{\mathrm{RTE}^{(i)}(n)\}_{i=1,2,\ldots,M}$, where $M = N/n$, and define the probability distribution of the finite-time RTE, $P(\mathrm{RTE}(n))$, by computing a frequency histogram of $\{\mathrm{RTE}^{(i)}(n)\}$, such the one shown in Figure \ref{fig:ftwve}(a) for $N = 10^{10}$ and $n = 200$. The inset in Figure \ref{fig:ftwve}(a) shows the finite-time RTE ``time series'' for the interval from $i = 40000$ to $i = 70000$. We see abrupt changes in the value of $\mathrm{RTE}(200)$, indicating the transitions from different regimes in the dynamics of the orbit. These changes in the value of the RTE cause its probability distribution to split into more than one mode. Szezech \textit{et al.} \cite{ftle} reported that for this value of $k$, the distribution of the finite-time Lyapunov exponent is bimodal. What we see is that the minor peak of Figure 3 in Ref. \cite{ftle} consists, in fact, of multiple peaks, as suggested by Harle and Feudel \cite{HARLE2007130}.

\begin{table}[b]
    \centering
    \caption{Correlation between the $\lambda_{\mathrm{max}}$ and the $\mathrm{RTE}$ for the standard map \eqref{eq:stdmap}.}
    \label{tab:corr}
    \begin{ruledtabular}
        \begin{tabular}{cc}
            Figure & $\rho_{\lambda_{\mathrm{max}}, \mathrm{RTE}}$\\
            \midrule
            \ref{fig:vs_k}(a) and \ref{fig:vs_k}(b) & 0.95\\
            \ref{fig:CGBDs}(a) and \ref{fig:CGBDs}(d) & 0.93\\
            \ref{fig:CGBDs}(b) and \ref{fig:CGBDs}(e) & 0.89\\
            \ref{fig:CGBDs}(c) and \ref{fig:CGBDs}(f) & 0.94\\
        \end{tabular}
    \end{ruledtabular}
\end{table}
\begin{figure}[tb]
    \centering
    \includegraphics[width=0.95\linewidth]{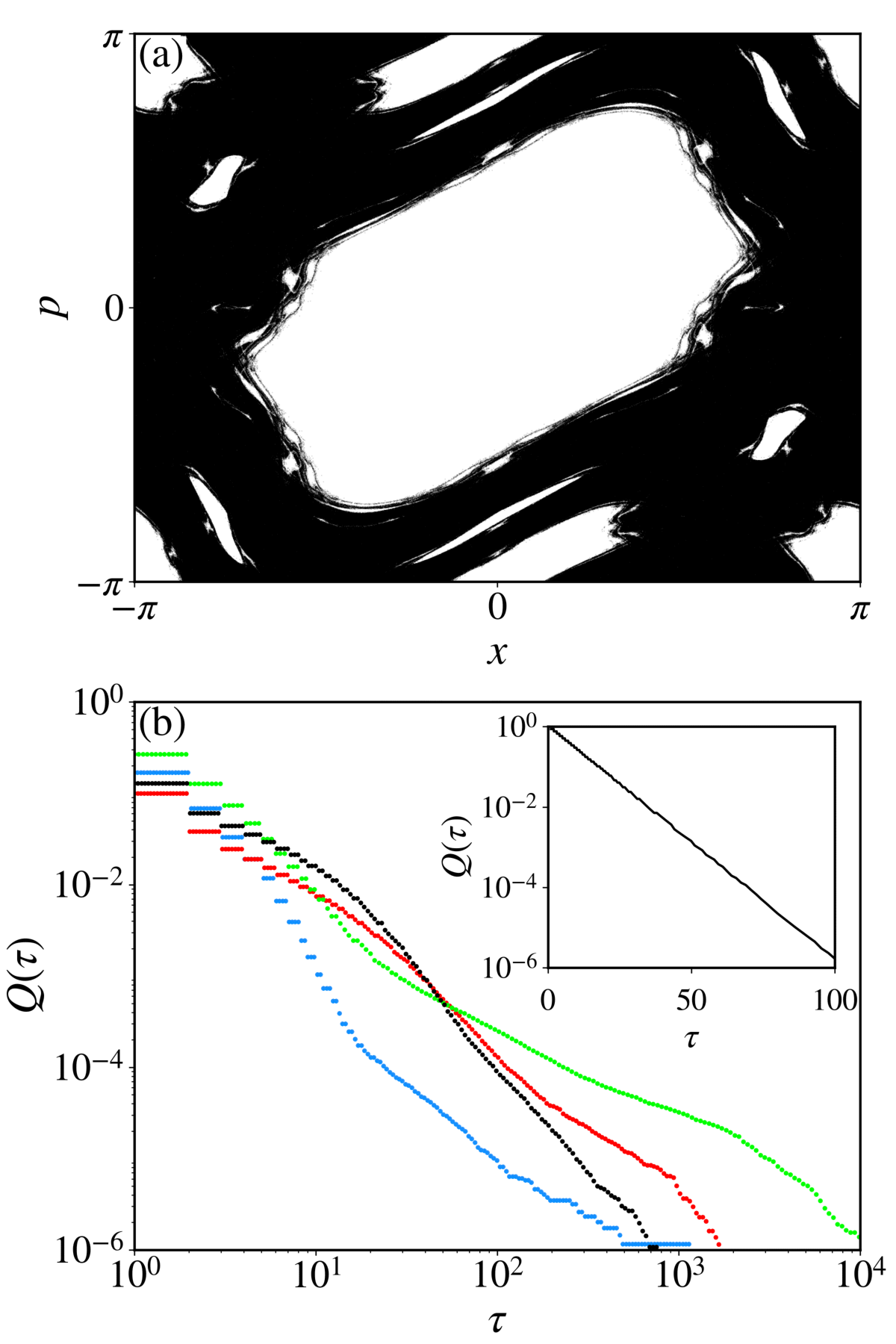}
    \caption{(a) The phase space points that generate the larger peak for high values of RTE in Figure \ref{fig:ftwve}(a) and (b) log-log plot of $Q(\tau)$ for each sticky region identified in Figure \ref{fig:ftwve}(a) with $N = 10^{11}$ and $n = 200$ (colored dots). Inset: Log-lin plot of $Q(\tau)$ of the phase space points shown in (a). The colors of the dots in (b) correspond to the colors of Figure \ref{fig:ftwve}.}
    \label{fig:qtau}
\end{figure}




The multi-modal distribution is due to the infinite hierarchical islands-around-islands structure embedded in the phase space. When the orbit is in the chaotic sea, the time-$n$ RTE is high, corresponding to the largest maximum of the distribution. On the other hand, when the orbit is trapped near an island, the RTE is low and the distribution exhibits smaller maxima for small values of $\mathrm{RTE}(200)$. Once trapped in the neighborhood of an island, the orbit may enter an inner level in the hierarchical structure and these transitions to different levels are the cause of the multi-modal distribution \cite{HARLE2007130}. Moreover, the closer to zero is $\lambda_{\mathrm{max}}$, the higher the hierarchical level of the island on which neighborhood the orbit is trapped \cite{Zaslavsky_hierarchy}. Hence multiple peaks are formed for small values of RTE [Figure \ref{fig:ftwve}(a)].

In order to identify the regions in phase space that correspond to the peaks in the distribution, we monitor the $\mathrm{RTE}(200)$ time series and plot the $200$ phase space positions $(x, p)$ with different colors for different ranges of $\mathrm{RTE}(200)$. The blue, red, green, and black points represent the phase space position when $\mathrm{RTE}(200) \in [0.6, 0.9]$, $\mathrm{RTE}(200) \in [1.2, 1.311]$, $\mathrm{RTE}(200) \in (1.311, 1.513]$, and $\mathrm{RTE}(200) \in (1.513, 1.66]$, respectively [Figures \ref{fig:ftwve}(b) and \ref{fig:ftwve}(c)]. Different peaks are indeed related to different hierarchical levels of the structure and by means of the RTE it is possible to distinguish between them very well. Additionally, the black and green points shadow the manifolds along which the non-trapped orbits leave the sticky region \cite{sticki_suppress}. And even though only the corresponding hierarchical levels around the period-6 island are shown [Figure \ref{fig:ftwve}(c)], all island chains contribute to the finite-time RTE distribution. For completeness, we also investigate the phase space points that generate the peak for high values of RTE. Whenever $\mathrm{RTE}(200) \in [2.5, 4.0]$ we plot the $200$ phase space points $(x, p)$ [Figure \ref{fig:qtau}(a)]. The phase space components in Figures \ref{fig:ftwve}(b) and \ref{fig:qtau}(a) complement each other: in Figure \ref{fig:qtau}(a) the chaotic sea is shown, \textit{i.e.}, the hyperbolic component of phase space, whereas in Figure \ref{fig:ftwve}(b) we see the nonhyperbolic component. Nonhyperbolicity can inhibit chaotic orbits from visiting some regions due to tangencies between stable and unstable manifolds \cite{nonhyperbolicity1, nonhyperbolicity2,nonhyperbolicity3}.

We can also measure the ``trapping time'' $t$ spent in each of the stickiness regimes, \textit{i.e.}, the time between two consecutive abrupt changes in the RTE. In Figure \ref{fig:ftwve}(a) we observe very clearly these trappings and we consider the boundary of each peak, defined by the filling colors, as the limits of the stickiness regimes. With the $\mathrm{RTE}(200)$ time series we obtain a set of trapping times $\{t_j\}_{j=1,2,\ldots,N_t}$ and define the probability distribution of the trapping times $P(t)$. Alternatively, we define the cumulative distribution of the trapping times as
\begin{equation}
    Q(\tau) = \sum_{t > \tau}P(t) = \frac{N_{\tau}}{N_t},
\end{equation}
where $N_{\tau}$ is the number of trapping times $t > \tau$ and $N_t$ the total number of them. It is well established in the literature that the distribution of the trapping times (and also its cumulative distribution) for fully chaotic systems has an exponential decay, whereas for quasi-integrable Hamiltonian systems which exhibit the stickiness effect, the decay obeys a power-law \cite{GRASSBERGER1985167,meiss_transport,PhysRevLett.82.528,ZASLAVSKY2002461,PhysRevE.67.046209,PhysRevLett.100.184101, crt1, crt2}. Using the finite-time RTE we are indeed able to separate these two different behaviors present in the dynamics, namely, the hyperbolic and nonhyperbolic ones. The cumulative distribution of the trapping times of the hyperbolic region is indeed exponential [inset of Figure \ref{fig:qtau}(b)], while $Q(\tau)$ of the nonhyperbolic regions have a power-law tail for large times [colored dots of Figure \ref{fig:qtau}(b)]. 


\section{Conclusion}
\label{sec:conclusion}

Several approaches to detect the existence of sticky orbits in the phase space of two-dimensional area-preserving systems have already been proposed and studied in previous studies. RP-based measures have also been used for this purpose, however without as much care as the Lyapunov exponents, for example. In this paper we have proposed the use of the Shannon entropy of the distribution of the recurrence times, RTE, estimated from the recurrence plots, to detect and characterize the stickiness effect in the standard map. We have shown that the RTE is positively correlated to the largest Lyapunov exponent, with a high correlation coefficient, and that it is possible to distinguish among the different types of motion using this entropy-based measure.

It is well-known that for systems that exhibit the stickiness effect, the transitions from fully chaotic motion to different levels in the hierarchical structure of islands-around-islands cause the FTLE distribution to have more than one peak. In fact, the distribution is multi-modal, in which each peak represents a different hierarchical level of islands. For the chosen nonlinearity parameter, $k = 1.5$, it was previously reported that the FTLE is bimodal \cite{ftle}. However, we have shown here that the maximum for small values of $\lambda_{\mathrm{max}}$/$\mathrm{RTE}$ is, in fact, composed of several minor peaks, as suggested by Harle and Feudal \cite{HARLE2007130}. This suggests that the RTE can be an alternative way of characterizing the stickiness effect, with which it is possible to distinguish among the different hierarchical levels in the islands-around-islands structure embedded in the chaotic component of phase space.

The presence of sticky regions in phase space affects global properties of the system, such as the distribution of the trapping times. We have shown that, by monitoring the finite-time RTE time series, it is possible to collect a set of trapping times $\{t_j\}$ for each of the different hierarchical levels detected by the finite-time RTE distribution and obtain the probability distribution of $\{t_j\}$, $P(t)$, and its cumulative distribution, $Q(\tau)$, of each hierarchical level. $Q(\tau)$ of fully chaotic systems has an exponential decay, whereas $Q(\tau)$ exhibits a power-law tail when sticky regions are present in the phase space. After separating among the distinct hierarchical levels, we have shown that the cumulative distribution of the hyperbolic component has indeed an exponential decay and that the power-law tail, characteristic of stickiness, is observed in the distribution of the different hierarchical levels.

One interesting point we plan to investigate in the future is whether the RTE can characterize also higher dimensional systems (\textit{e.g.} 4D symplectic maps) using the methodology presented in this paper. Another interesting study is the critical value $k = 0.971635$\cite{greene}, where the last invariant rotational torus ceases to exist. For this parameter, there is a sequence of \textit{cantori} and this could give an interesting histogram for the finite-time RTE.

All of our simulations regarding the evaluation of the recurrence matrix were made using the \textit{pyunicorn} package \cite{pyunicorn} and even though these results were obtained to the standard map, we expect similar results for any quasi-integrable Hamiltonian system that exhibit stickiness.

\begin{acknowledgments}
We wish to acknowledge the support of the Araucária Foundation, the Coordination of Superior Level Staff Improvement (CAPES), under Grant No. 88881.143103/2017-1, the National Council for Scientific and Technological Development (CNPq), under Grant Nos. 403120/2021-7, 301019/2019-3 and São Paulo Research Foundation (FAPESP) under Grant Nos. 2018/03211–6, 2022/04251-7. J.K. has been supported by the Alexander von Humboldt Polish Honorary Research Scholarship 2020 of the Fundation for Polish Science. We would also like to thank the 105 Group Science \footnote{\url{www.105groupscience.com}} for fruitful discussions.
\end{acknowledgments}

\section*{Code availability} 

The source code to reproduce the results of this paper is freely available in the Zenodo archive \cite{code}.
    
\section*{Declaration of competing interest}

The authors declare that they have no known competing financial interests or personal relationships that could have appeared to influence the work reported in this paper.

\appendix

\section{The effect of the threshold on the RTE}
\label{sec:appendix}

In Section \ref{sec:recplot} we introduced the concept of RPs and made some considerations regarding the choice of the threshold $\epsilon$: we chose $\epsilon$ to be $10\%$ of the time series standard deviation, $\sigma$. In the following, we provide an analysis of the effect of $\epsilon$ in our results.

\begin{figure}[t]
    \centering
    \includegraphics[width=0.9\linewidth]{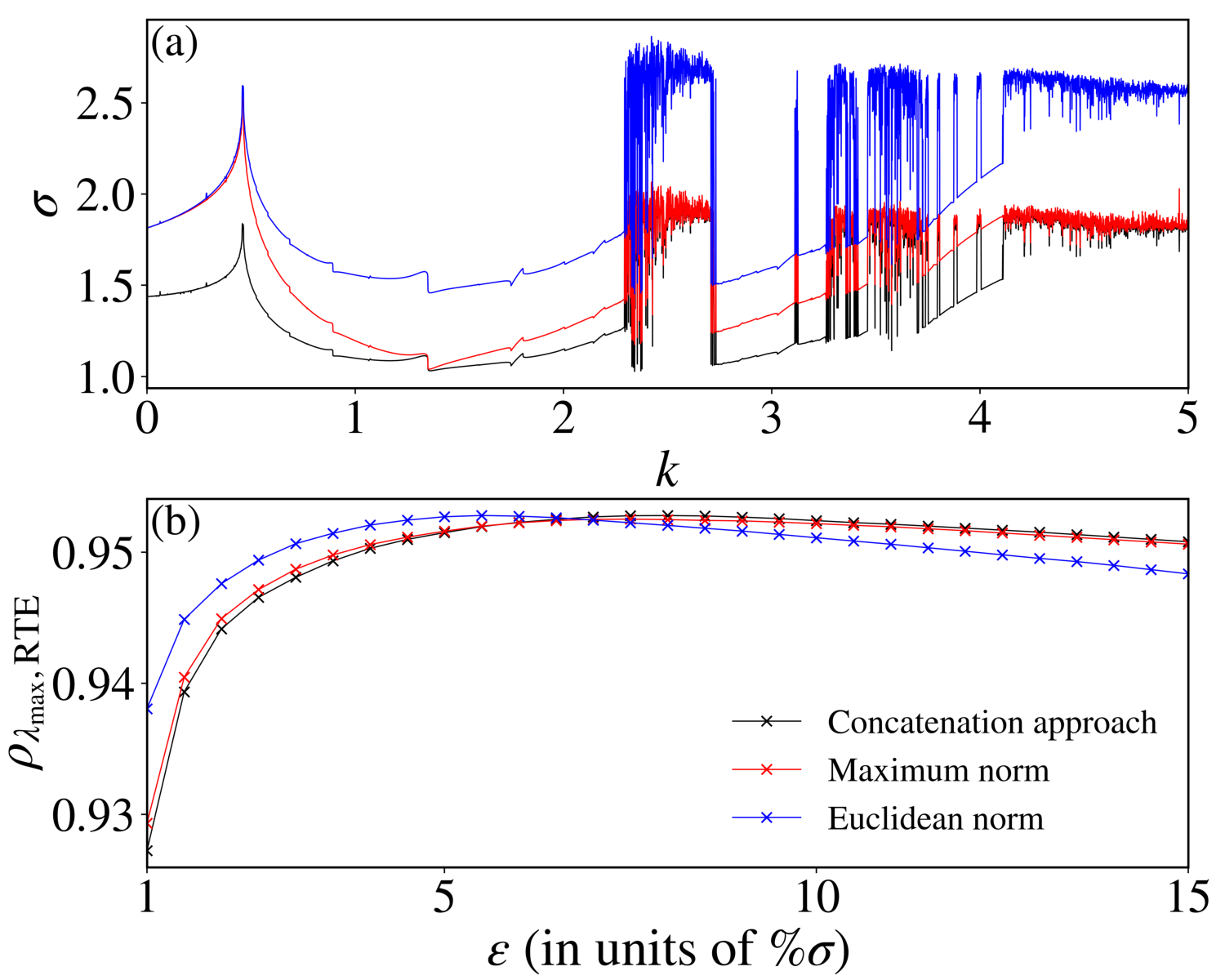}
    \caption{(a) The standard deviation, $\sigma$, as a function of $k$ for the orbit with initial condition $x_0 = 0.0$ and $p_0 = 1.3$ using the concatenation approach (black) and the norm approach (red and blue), considering the maximum and Euclidean norms, respectively, and (b) the correlation coefficient between $\lambda_{\mathrm{max}}$ and RTE as a function of the threshold $\epsilon$ (in units of the percentage of $\sigma$).}
    \label{fig:figA1}
\end{figure}

When dealing with $d$-dimensional data, the problem of how to calculate its standard deviation arises. The simplest approach one could consider is to concatenate the time series of each component, creating a new $dN$-dimensional vector $(x^{(1)}_1, x^{(1)}_2, \ldots, x^{(1)}_N, x^{(2)}_1, x^{(2)}_2, \ldots, x^{(2)}_N, \ldots, x^{(d)}_1, x^{(d)}_2, \ldots, x^{(d)}_N)^T$ ($N$ is the time series length), and compute its standard deviation. Another approach one could choose is to consider a standard deviation vector, $\vec{\sigma}$, where each component is the standard deviation of each time series individually, and compute its norm. Here we consider the maximum and the Euclidean norms, given by
\begin{subequations}
    \begin{align}
        \norm{\vec{\sigma}}_{\infty} &= \mathrm{max}(\sigma_1, \sigma_2, \ldots, \sigma_d),\label{eq:norminf}\\
        \norm{\vec{\sigma}}_2 &= \sqrt{\sigma_1^2 + \sigma_2^2 + \ldots + \sigma_d^2},\label{eq:norm2}
    \end{align}
\end{subequations}
respectively. Using the maximum norm corresponds to choosing the maximum of all standard deviations of the different time series (different components of $\vec{\sigma}$), while the Euclidean norm corresponds to the ordinary distance from the origin to the point $\vec{\sigma}$ in the ``standard deviation space''. Figure \ref{fig:figA1}(a) shows the standard deviation calculated using the concatenation approach (black) and the norm approach, considering the maximum (red) and the Euclidean (blue) norms, as a function of $k$, with the same initial condition as in Figure \ref{fig:vs_k}. The concatenation and norm (maximum) approaches, yield similar standard deviations, while the norm approach with Euclidean norm yields a larger value of $\sigma$. However, all methods agree that chaotic orbits have a larger standard deviation.

To determine the optimal method for calculating $\sigma$ and an appropriate value of $\epsilon$, we compute the correlation coefficient, Eq. \eqref{eq:correlationcoef}, between $\lambda_{\mathrm{max}}$ and RTE as a function of $\epsilon$ (in units of $\%\sigma$) using the concatenation and norm approaches to calculate $\sigma$ [Figure \ref{fig:figA1}(b)]. The three approaches yield similar correlation coefficients. Even if we choose a very small $\epsilon$ ($1\%$ of $\sigma$), we still obtain a high correlation coefficient ($\approx 0.93$). Also, increasing $\epsilon$ does not affect significantly $\rho_{\lambda_{\mathrm{max}}, \mathrm{RTE}}$, indicating that in our case the choice of $\epsilon$ is not that sensible as it would be in other cases. In fact, there is a range of values for $\epsilon$ in which the results are good.

Even though the three approaches of calculating the standard deviation give similar results in our case, choosing the concatenation approach when one time series has a different value range than the others might strongly bias the standard deviation. Therefore, in our opinion, the most appropriate approach is the norm approach (either Euclidean or maximum). 

%

\end{document}